\documentclass[12pt]{iopart}

\usepackage{iopams}
\usepackage{xfrac}
\usepackage{graphicx}
\usepackage{epstopdf}
\usepackage{float}

\usepackage{caption}
\usepackage{subcaption}

\newcommand{\be}{\begin{equation}}
\newcommand{\ee}{\end{equation}}
\newcommand{\bmat}{\begin{bmatrix}}
\newcommand{\emat}{\end{bmatrix}}
\newcommand{\bfg}{\begin{figure}[h]\centering}
\newcommand{\efg}{\end{figure}}
\newcommand{\ba}{\begin{eqnarray}}
\newcommand{\ea}{\end{eqnarray}}
\newcommand{\dpp}{\partial}

\newcommand{\hml}{\mathcal{H}}

\newcommand{\la}{\langle}
\newcommand{\ra}{\rangle}
\newcommand{\PP}{\mathbb{P}}

\begin{document}

\title{Entanglement, squeezing and state-swap in cholesteric liquid crystals}

\author{J D P Machado, Ya M Blanter}

\address{Kavli Institute of Nanoscience, Delft University of Technology, Lorentzweg 1, 2628 CJ Delft, The Netherlands}
\vspace{10pt}
\begin{indented}
\item[]\date{\today}
\end{indented}

\begin{abstract}
Phenomena such as state-swap, quadrature squeezing, entanglement and violation of entanglement inequalities are frequently reported to occur in quantum systems only. It is shown here that these effects can also occur in cholesteric liquid crystals in the presence of an applied magnetic field. The nature of these effects as well as the conditions for their observation are analysed.
\end{abstract}

\maketitle

\section{Introduction}

The question regarding the quantum nature of physical phenomena is often incorrectly deemed as a linguistic issue. As theories are often considered simply as a suitable language insofar as they are able to model experimental observations, what differentiates a semantic quarrel from a relevant physical problem when classifying potentially quantum effects is the consequences of such classification. If certain phenomena are not inherently quantum, they could be observed for a wider range of systems outside the quantum realm. Currently, the most reliable way to cast a verdict in the quantum-classical comparison is with Bell-like inequalities \cite{Belllike}, as the violation of these inequalities cannot take place in classical systems. However, Bell inequalities are not able to characterise the nature of an effect, and it is believed that phenomena such as squeezing, entanglement and state-swap are intrinsically quantum \cite{palomakias,squeezu,alemaes,fotosqueezing,charmichael,riviu,italianos} even when no Bell-like inequality is violated. With the observation of these effects in increasingly larger and heavier systems \cite{sokawai,amiradosarabesunidos,squeezu,palomakias,statetransfer}, it is essential to understand the conditions for such effects to occur at a macroscopic scale. But their emergence at the macroscopic scale, and the fact that they are not inherently nonclassical, poses a question which has rarely been addressed: Is it possible to observe phenomena from the quantum realm in macroscopic systems without a quantum nature?

In this manuscript, we answer this question affirmatively. Resorting to a statistical formalism presented in Sec.\ref{sec:formalismo}, we show that entanglement, squeezing, and state-swap can occur in liquid crystals. Specifically, by analysing the statistical properties of the director field of a cholesteric liquid crystal in the presence of an applied magnetic field, we show how these effects relate to the fluctuations and distribution of the director. It is known that the (optical) properties of nematic liquid crystals are strongly determined by the fluctuations of the director \cite{flutuacoesLC1,flutuacoesLC2,flutuacoesLC3,flutuacoesLC4}, but the role of fluctuations of the director field has never been linked to the effects of quantum fluctuations in analogous systems. Here, we discuss the close connexion between the two, and propose a tangible  system where the aforementioned effects can be observed.

State-swap is a method to create a quantum state for a macroscopic object based on swapping its state with the state of another quantum element. State-swap between microwaves and a mechanical resonator has already been achieved for Gaussian states \cite{statetransfer}, entering the list of quantum effects \cite{riviu}. In Sec.\ref{sec:stateswap}, we show that state-swap also occurs in liquid crystals in the form of a swap between the fluctuations in the polar and azimuthal angles of the director field.

The inherent uncertainties of quantum mechanics impose natural limits on the precision of measurements, such as the zero-point motion (ZPM) for position measurements. These quantum restrictions inspired strategies for manipulating noise and reduce the imprecision beyond the standard quantum limit. For position measurements, the position uncertainty can be reduced below ZPM by transferring part of the uncertainty to the momentum. Squeezing refers to this uncertainty trade-off between conjugate variables, and it has been considered as a truly quantum \cite{alemaes} and nonclassical effect \cite{squeezu,fotosqueezing,charmichael}. Nevertheless, we show in Sec.\ref{sec:squeezing} that this is not a requirement to observe squeezing beyond the ZPM level, and we evaluate the degree of squeezing of the director field. Note that the squeezing effect discussed here could be related to the known reduction of director fluctuations by the application of AC electric fields \cite{flutuacoesLC4}.

Entanglement is the crown jewel of quantum mechanics and it has already been observed in $\mu$m-sized mechanical elements \cite{palomakias}. Though it refers only to the property that the measurement of a system determines the state of a second (sub)system, the lack of precedents in classical theories has lead entanglement to be deemed a peculiarity of the quantum world \cite{italianos}. Recent investigations showed that entanglement can also occur in classical systems \cite{brownianentanglement,epistemologia}, but no concrete proposals to observe entanglement in real physical systems were put forward so far. We show in Sec.\ref{sec:entanglement} that it is not only possible to observe entanglement in liquid crystals, but also that entanglement inequalities can be violated.

Finally, we evaluate the validity of our model and the feasibility of our proposal to observe these effects in Sec.\ref{sec:validade}.

\section{Statistical framework}
\label{sec:formalismo}

To model statistical behaviour in a dynamical system, the phase space must be dressed with a probability distribution $\PP$, and random variables must be assigned to the classical degrees of freedom. Two equivalent descriptions are possible: the random variables evolve in time just as the classical degrees of freedom would, and expected values are computed with a static probability distribution for the random variables; or the probability distribution evolves in time while the random variables remain static. These two possible descriptions are respectively the classical analog of the Heisenberg and Schr\"{o}dinger pictures in quantum mechanics. Based on this connexion, and on the fact that $\PP$ contains all the statistical information about the system's properties, we shall refer to $\PP$ as the state of the system, in analogy with the quantum quasiprobability distributions such as the $Q$-, Wigner and $\mathcal{P}$-functions. In Hamiltonian mechanics, the Poisson bracket $\{.,.\}$ governs the dynamics, and the time-evolution of the average value of a random variable $A$ is
\be
\fl \la d_tA(Q,P)\ra =\int dQdP\, \{A(Q,P),\hml\} \PP(Q,P)=-\int dQdP\,A(Q,P) \{\PP(Q,P),\hml\}\,,
\ee
with $(Q,P)$ being the pair of canonical variables and their conjugates. For the two descriptions to be equivalent for any random variable $A$, the time-evolution for $\PP$ must obey $d_t\PP=\{\hml,\PP\}$. In \ref{app:provas}, this equation is shown to preserve the normalisation and positivity of $\PP$.

Complementary to a Hamiltonian formulation, a Lagrangian description can also be employed. The Hamiltonian of a system is related to the Lagrangian $\mathcal{L}$ by $\hml=\dot{Q}P-\mathcal{L}$, where the canonical momentum $P$ is given by $P=\frac{\dpp \mathcal{L}}{\dpp \dot{Q}}$. The advantage of the Lagrangian formulation is that the system's dynamics can be derived from the minimisation of a functional (the action). The philosophy behind the principle of least action has analogues in other field theory formulations, namely, in continuum theories of soft condensed matter systems, where the equilibrium configuration of a field $\phi$ is given by the minimum of the free energy $F (\phi)=\int d\vec{r} \mathcal{F}(\phi,\nabla \phi)$. In this analogy, the spatial dimensions play the role of time, the free energy density $\mathcal{F}$ plays the role of the Lagrangian density, and the system's initial conditions correspond to the field distribution at the boundaries. Due to this similarity, the spatial statistical equilibrium properties of fields can be investigated in a similar manner to the time-evolution of the statistical properties of classical systems.

For simplicity, consider that the system's field degrees of freedom vary only along a direction $z$. With the mapping $t\to z$, $Q\to\phi$ and $\mathcal{L}\to \mathcal{F}$, the evolution equation given by the Poisson brackets becomes
\be
d_z\PP=\{\PP,\mathcal{F}\}+(\dpp_z\phi)\dpp_\phi P_\phi\frac{\dpp \PP}{\dpp P_\phi}-(\dpp_z\phi+P_\phi\dpp_{P_\phi})(\dpp_z\phi))\frac{\dpp \PP}{\dpp \phi}\,,
\label{eq:poissona}
\ee
where $P_\phi$ is the field's momentum ${\Large P_\phi=\frac{\dpp \mathcal{F}}{\dpp (\dpp_z\phi)} }$.

\section{Model}
\label{sec:modelo}

\begin{figure}[h]
\centering
\includegraphics[scale=0.6, angle=0]{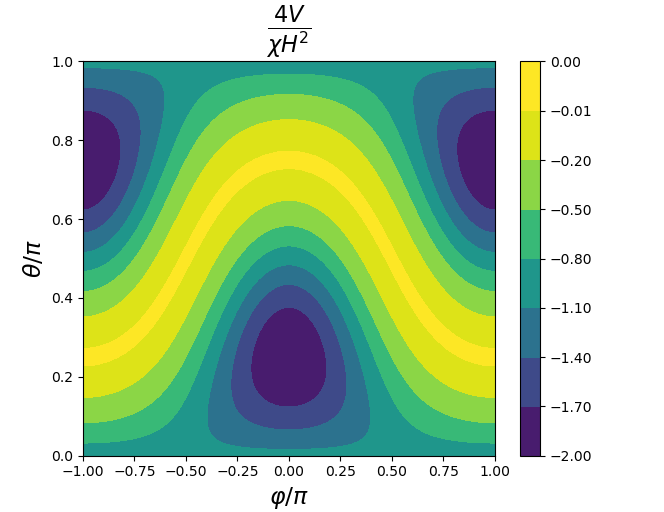}
\captionof{figure}{Potential created by the magnetic field as a function of $(\varphi,\theta)$. The potential has minima at $(\varphi,\theta)=(0,\pi/4)$ and $(\varphi,\theta)=(\pi,3\pi/4)$. For centro-symmetric molecules, both minima correspond to the same physical orientation as the ``front'' and ``back'' of a molecule are indistinguishable.}
\label{fig:potencial}
\end{figure}

The system considered is a cholesteric liquid crystal in the presence of a magnetic field $\vec{H}$. The free energy density $\mathcal{F}$ is given by \cite{deGennes,lavrentovich}
\be
\fl
\mathcal{F}=\frac{K_1}{2}(\nabla .\vec{n})^2+\frac{K_2}{2}\left(\vec{n}.(\nabla \times\vec{n})+\frac{2\pi}{P_0}\right)^2+\frac{K_3}{2}(\vec{n}\times(\nabla \times\vec{n}))^2-\frac{\chi_a}{2}(\vec{H}.\vec{n})^2\,,
\label{eq:energialivre}
\ee
where $\vec{n}=\cos\varphi\sin\theta \vec{e}_x+\sin\varphi\sin\theta \vec{e}_y+\cos\theta \vec{e}_z$ is the director field, $K_1$, $K_2$ and $K_3$ are the splay, twist and bend elastic moduli, $P_0$ is the pitch of the cholesteric helix, and $\chi_a$ is the difference between the parallel and perpendicular magnetic susceptibilities. Though we only consider an applied magnetic field,  the same physics occurs for an applied electric field. By applying the field $\vec{H}=H/\sqrt{2}(\vec{e}_x+\vec{e}_z$), one creates the potential
\be
V=-\frac{\chi_a}{4}H^2(\cos \varphi \sin \theta + \cos \theta )^2\,,
\ee
which has a minimum at $(\theta,\varphi)=(\pi/4,0)$  for $\chi_a>0$ (the usual case \cite{deGennes}) as seen in Fig. \ref{fig:potencial}. When $\chi_a>0$, it costs energy for the director to misalign from the magnetic field direction, and the potential minimum is fixed by the orientation of the magnetic field. The choice $\vec{H}\propto \vec{e}_x+\vec{e}_z$ is to maximise the  effect of the cholesteric pitch.

When the director is close to the potential minimum, Eq.(\ref{eq:energialivre}) can be linearised around the minimum, and to further enhance the effects of the pitch, the mode-matching condition $K_1=2K_2$ is considered. Additionally, 
we consider the case where a wall in the $x-y$ plane fixes the orientation of the director at $z=0$, and assume the director is only able to vary in the direction perpendicular to the wall (the $z$-direction), and focus our attention in the competition between the effects of the magnetic field in the bulk and the prescribed director at the boundary. With these assumptions, and using the dimensionless variables $q_\theta=\theta-\pi/4$, $q_\varphi=\varphi-0$, $\zeta=\lambda z$, $\lambda=H\sqrt{\chi_a /(K_2+K_3)}$, and $\eta=4\pi \lambda K_2 /(P_0\chi_a H^2)$, the free energy Eq.(\ref{eq:energialivre}) becomes 

\be
\mathcal{F}=\varepsilon_m\left[2(\dpp_\zeta q_\theta)^2 + (\dpp_\zeta q_\varphi)^2 + 2q_\theta^2 + q_\varphi^2 + 2\eta(\dpp_\zeta q_\varphi)(1+2q_\theta)\right]\,,
\label{eq:Fquad}
\ee
with $\varepsilon_m=\chi_aH^2/4$. The equation above states that, without a cholesteric pitch, the elastic energy must compete with the magnetic field, leading to spatial variations of the director. The director changes orientation along $z$ to align with the anchoring direction at the boundary and with the magnetic field  in the bulk, with the change occurring over a typical length of $\lambda^{-1}$. Furthermore, the azimuthal and polar angles vary independently. The effect of the cholesteric pitch is to couple both degrees of freedom, and it can thus be interpreted as an interaction between the polar and azimuthal angles.

Intuitively the director can be seen as the average molecular orientation over volumes much larger than molecular sizes, but smaller than typical deformation lengths. As the molecules are not fully aligned locally, deviations from the average orientation can be seen as fluctuations in the director field, which can occur, for example, due to thermodynamical fluctuations.

The continuum theory exposed above is based on treating the director field as a local average of the molecular orientation. The orientation of each molecule has deviations from the director (and the orientation itself may also fluctuate), and these deviations can be interpreted as fluctuations of the director field and treated statistically. The director fluctuations can be taken into account by extending the formalism by assigning a probability distribution $\PP$ to the director's orientation. The spatial distribution of $\PP$ is obtained by minimising the free energy as discussed in Sec.\ref{sec:formalismo} (c.f. Eq.(\ref{eq:poissona})).

To find and model the spacial evolution of the director probability distribution, it is handier to use the dimensionless momenta
\be
\pi_\theta=\frac{1}{\varepsilon_m}\frac{\dpp \mathcal{F}}{\dpp (\dpp_\zeta q_\theta)}=4\dpp_\zeta q_\theta\,,
\label{eq:pi_teta}
\ee
\be
\pi_\varphi=\frac{1}{\varepsilon_m}\frac{\dpp \mathcal{F}}{\dpp (\dpp_\zeta q_\varphi)}=2(\dpp_\zeta q_\varphi + \eta (1+2q_\theta))\,.
\label{eq:pi_fi}
\ee

Using these momenta and Eq.(\ref{eq:poissona}), the spatial-evolution of $\PP$ is given by the equation
\be
\fl
\dpp_\zeta \PP= - \left(4(1-2\eta^2)q_\theta +2 \eta\pi_\varphi - 4\eta^2 \right)\frac{\dpp \PP}{\dpp \pi_\theta}-\frac{1}{4}\pi_\theta\frac{\dpp \PP}{\dpp q_\theta}-2q_\varphi\frac{\dpp \PP}{\dpp \pi_\varphi} - \left(\frac{\pi_\varphi}{2} -(1+2 \eta q_\theta)\right) \frac{\dpp \PP}{\dpp q_\varphi}\,.
\label{eq:stocastica}
\ee

The solution to this linear first order partial differential equation is straightforward to find with the method of characteristics, and details are presented in \ref{app:dinamica}. From here on, only the strong-coupling limit $\eta\gg\sqrt{2}$ shall be considered because the spatial-evolution simplifies to purely oscillatory behaviour. Integrating in the momenta $\pi_{\theta,\varphi}$, the marginal probability distribution for the director angles is given by
\be
\fl
\PP(\zeta)=|\xi|\,\PP_\varphi\Big(\frac{\cos(\sqrt{2}\eta\zeta)}{\xi}q_\varphi+\frac{\sqrt{2}\sin(\sqrt{2}\eta\zeta)}{\xi}q_\theta\Big)\PP_\theta\left(-\frac{\sin\left(\zeta/(\sqrt{2}\eta)\right)}{\sqrt{2}\xi}q_\varphi+\frac{\cos\left(\zeta/(\sqrt{2}\eta)\right)}{\xi}q_\theta\right)\,,
\label{eq:marg_prob}
\ee
where $\PP_{\theta,\varphi}$ are respectively the probability distributions for the angles $\{\theta,\varphi\}$ at the boundary $\zeta=0$ and 
\be
\xi = \cos \left(\frac{\zeta}{\sqrt{2}\eta}\right) \cos(\sqrt{2}\eta\zeta) + \sin\left(\frac{\zeta}{\sqrt{2}\eta}\right) \sin(\sqrt{2}\eta\zeta)\,.
\ee

\section{State-swap}
\label{sec:stateswap}

Consider that the probability distribution for the director angles at the boundary $\zeta=0$ factorises as $\PP(\zeta=0)=\PP_\theta(q_\theta) \PP_\varphi(q_\varphi)$. Using Eq. (\ref{eq:marg_prob}), when $\eta=\sqrt{\frac{1+4m}{6+8n}}$ (with $m,n$ integers obeying $m\gg 3+4n$ and $n>0$), the angular probability distribution at $\zeta_T=\sqrt{8}\eta\pi (n+3/4)$, becomes
\be
\PP(\zeta_T)=\PP_\varphi\Big(\sqrt{2}q_\theta\Big)\PP_\theta\left(\frac{1}{\sqrt{2}}q_\varphi \right)
\label{eq:state-transfer}
\ee

\begin{figure}[h]
\fl
\centering
    \begin{subfigure}[b]{0.39\textwidth}
        \includegraphics[width=\textwidth]{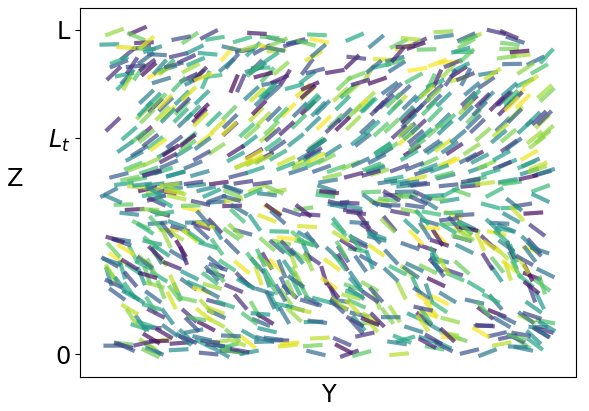}
        \caption{}
        \label{fig:st1}
    \end{subfigure}
    \begin{subfigure}[b]{0.36\textwidth}
        \includegraphics[width=\textwidth]{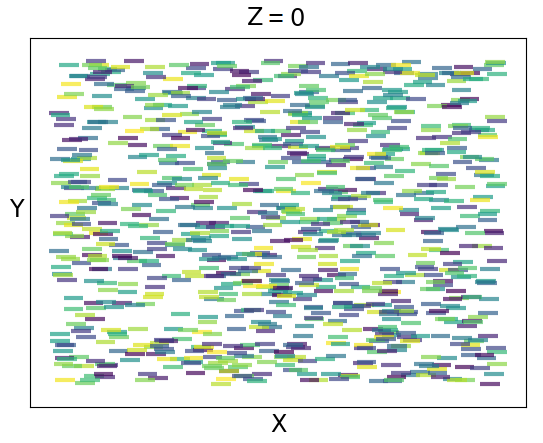}
        \caption{}
        \label{fig:st2}
    \end{subfigure}
    \begin{subfigure}[b]{0.36\textwidth}
        \includegraphics[width=\textwidth]{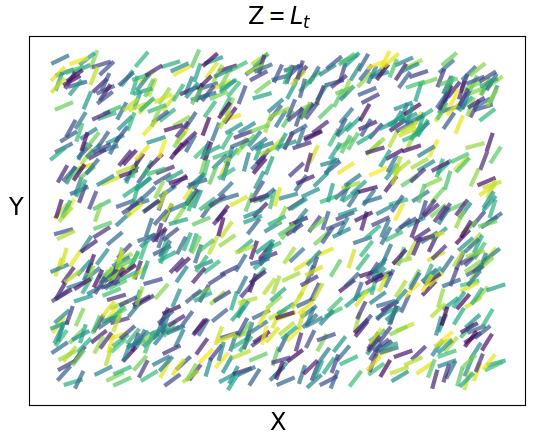}
        \caption{}
        \label{fig:st3}
    \end{subfigure}
    \caption{Illustration of state-swap in liquid crystals. The molecules at $z=0$ display deviations from the director in the orientation along $z$ (panel \ref{fig:st1}), but no deviations in the perpendicular plane (panel \ref{fig:st2}). In contrast, the molecules display no deviations from the director at $L_T=\zeta_T/\lambda$, but display the same amount of deviation from the director in the perpendicular plane (panel \ref{fig:st3}) as the deviations along $z$ at $z=0$. This transfer of distributions of deviations is what constitutes state-swap.}
\label{fig:unica}
\end{figure}

The physical meaning of Eq.(\ref{eq:state-transfer}), illustrated in Fig. \ref{fig:unica}, is that at $\zeta_T$, the probability distribution for the azimuthal angle has swapped with the one for the polar angle. The extra $\sqrt{2}$ factor is due to the potential minimum around which linearisation is performed and because the angles $\theta,\varphi$ have different span sizes ($[0,\pi]$ vs. $]-\pi,\pi]$). Apart from this factor, the statistics of any measurement of $\varphi$ at $\zeta_T$ will faithfully reproduce the statistics of $\theta$ at $\zeta=0$ and vice-versa. This effect is akin to the state-swap observed in electromechanics \cite{statetransfer}, and it shows the possibility of state-swap in systems without a quantum nature.

\section{Squeezing}
\label{sec:squeezing}

\begin{figure}[h]
\fl
\centering
    \begin{subfigure}[b]{0.57\textwidth}
        \includegraphics[width=\textwidth]{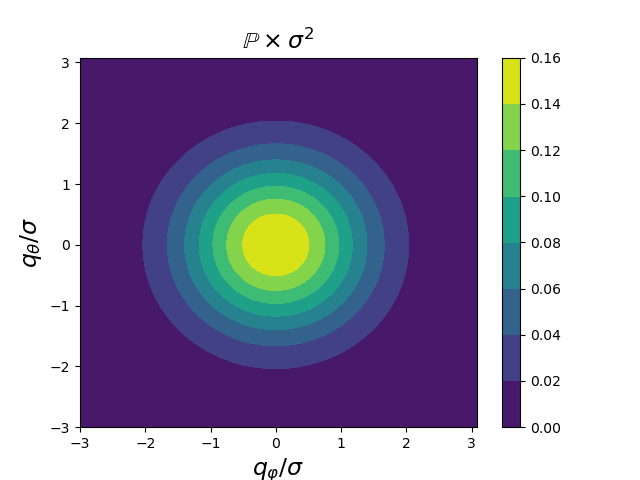}
        \caption{State at $\zeta=0$}
        \label{fig:Gaus_state_boundary}
    \end{subfigure}
     \begin{subfigure}[b]{0.57\textwidth}
        \includegraphics[width=\textwidth]{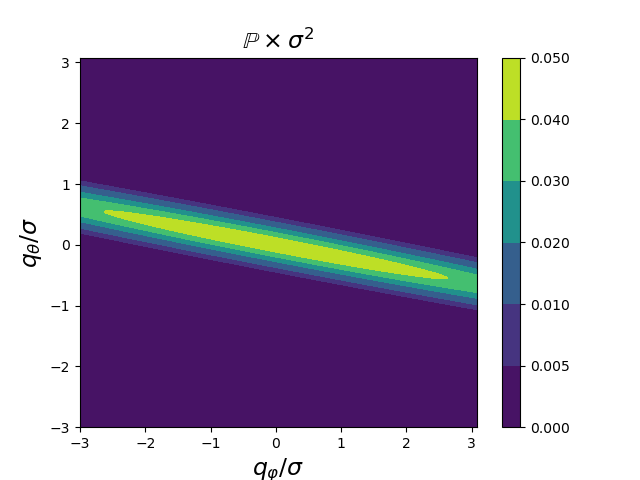}
        \caption{State at $\zeta_S$}
        \label{fig:Squee_state}
    \end{subfigure}
    \caption{Angular probability distributions at the boundary $\zeta=0$ and at $\zeta_S$, for $\eta=2.05$. The effect of squeezing is visible by the shrinkage of the variance of $q_\theta$, while enlarging the variance of $q_\varphi$, and it arises due to the interaction mediated by the cholesteric pitch.}
\label{fig:squeezed_state}
\end{figure}

Another consequence of the interaction in a statistical framework is the change in the quadratures' variance, which characterises the uncertainties associated with quadrature measurements. Consider that the probability distribution of the director angles at the boundary $\zeta=0$ is
\be
\PP(\zeta=0)=\frac{1}{2\pi\sigma^2}\exp\left(-\frac{1}{2\sigma^2}\left(q_\varphi^2+q_\theta^2\right)\right)\,.
\label{eq:prob_s0}
\ee
For simplicity, consider the layer at $\zeta_S = \sqrt{2}\eta \pi$. At $\zeta_S$, the probability distribution becomes
\be
\fl
\PP(\zeta_S)=\frac{|\cos(2\pi\eta^2)|}{2\pi\sigma^2}\exp\left(-\frac{1}{2\sigma^2}\left(q_\varphi^2+\left(1+2\tan^2(2\pi\eta^2)\right)q_\theta^2+\sqrt{8}\tan(2\pi\eta^2)q_\theta q_\varphi\right)\right)\,.
\label{eq:prob_squeezu}
\ee
This state is shown in Fig.\ref{fig:squeezed_state}, where it is visible that the rotationally symmetric Gaussian state at the boundary is squeezed along a given direction, and the variance of $q_\theta$ is reduced at the expense of enlarging the variance of $q_\varphi$. It is this reduction in a quadrature's variance, while enlarging its counterpart, that defines squeezing. The physical meaning of this uncertainty reduction in $q_\theta$ is that the cholesteric pitch forces the molecules to align better with the director along the $z-$direction, while loosening the alignment in the perpendicular plane.  By measuring the orientation of the molecules along $\zeta$, this effect should be visible from the sharp alignment of the molecules at $\zeta_S$.

\begin{figure}[h]
\centering
\includegraphics[scale=0.7, angle=0]{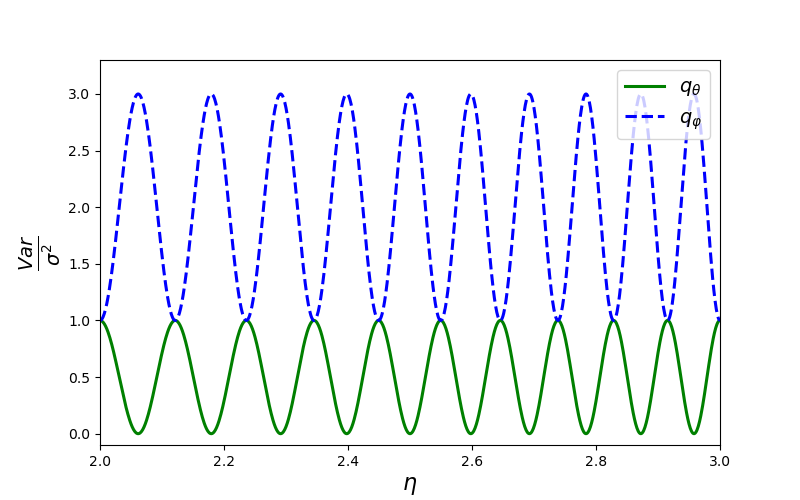}
\captionof{figure}{Dependence of the angular variance at $\zeta_S$ with the interaction parameter $\eta$.}
\label{fig:squeez_var}
\end{figure}

With Eq.(\ref{eq:prob_squeezu}), the variances can be found to be given by
\be
Var(q_\theta) = \sigma^2 \cos^2(2\pi\eta^2)\quad \text{and}\quad Var(q_\varphi) = \sigma^2 (1 + 2\sin^2(2\pi\eta^2))\,.
\ee
The quadrature variances possess an oscillatory dependence displayed in Fig. \ref{fig:squeez_var}. The reason for the $q_\theta$ variance to be squeezed while the $q_\varphi$ variance is enlarged is solely due to the choice made for $\zeta_S$. Other choices promote different squeezed quadratures and degrees of squeezing.

The uncertainty trade-off between quadratures portrayed here is analogous to the one measured experimentally in quantum optical contexts \cite{fotosqueezing,sokawai,amiradosarabesunidos,squeezu}). Thus, the statistical nature of liquid crystals allows for the observation of squeezing, and for linear couplings, the degree of squeezing is the same as in the respective quantum theory.

Although squeezing has been observed outside the quantum regime \cite{thermalsqueezing}, it is still believed that reducing the noise below ZPM signals the transition to the quantum realm. However, there are no limitations to classical squeezing, and the molecular orientation uncertainty can be reduced below ZPM, depleting the significance of such threshold. The connexion between squeezing and quantum/nonclassical criteria is based on the negativity of the $\mathcal{P}$-function \cite{charmichael}, but a negative $\mathcal{P}$-function does not necessarily display nonclassical features \cite{Diosi}, and the claim of a quantum nature for squeezing is not based on physical observables. Consequently, there is no physical impediment to observe a director's orientation variance below the zero-point uncertainty.

\section{Entanglement}
\label{sec:entanglement}

Entanglement refers to the property of a deterministic measurement outcome of A, once the random outcome of a previous measurement of B is known, with $A$ and $B$ observables of distinct elements. Though it often acquires unjustified spiritistic hues, its nature is a simple statistical property.

Consider that the anchoring forces at the boundary $\zeta=0$ fix the azimuthal angle to be sharply defined around $q_\theta=0$, and the polar angle to have an equiprobable distribution of range $2d$, centred around $q_\varphi=0$. For completeness, consider that the polar momentum $\tilde{\pi}_\varphi$ is sharply defined at the boundary whereas the probability distribution for $\pi_\theta$ is a Gaussian of variance $\sigma^2$ centred around the origin.  When $\eta=\sqrt{\frac{3+8m}{6+16n}}$ (with $m,n$ integers obeying $m\gg 4n$ and $n>0$), the probability distribution at $\zeta_E=\sqrt{8}\eta\pi (n+3/8)$ becomes (c.f. \ref{app:dinamica} for details on the momenta)

\ba
\fl
\PP(\zeta_E)=\frac{1 }{\sqrt{\pi} \sigma d} \,\mathbb{I} \left( \left| q_\theta - \frac{q_\varphi}{\sqrt{2}} \right|< d \right) \delta \left( q_\theta + \frac{q_\varphi}{\sqrt{2}} \right)\nonumber\\
\times \delta \left(\tilde{\pi}_\varphi - \frac{\pi_\theta}{\sqrt{2}}+\sqrt{8}\eta q_\varphi \right)
\exp \left( -\frac{(\tilde{\pi}_\varphi + \frac{\pi_\theta}{\sqrt{2}} + 4 \eta q_\theta)^2 }{2\sigma^2}\right)\,,
\label{eq:prob_entang}
\ea
where $\mathbb{I}$ is the indicator function. The state in Eq.(\ref{eq:prob_entang}) features strong correlations between the random variables, and these correlations can be characterised by the Pearson correlation coefficient $PCC(x,y)=\frac{CoVar(x,y)}{\sigma(x)\sigma(y)}$, where $\sigma(x)$ and $CoVar$ are respectively the standard deviation and the covariance. Here onwards, we consider the limit case where momentum fluctuations dominate over angular fluctuations, i.e.
$\beta = 4\eta d /(\sqrt{3}\sigma) \ll 1$.

The $PCC$ matrix for the probability distribution in Eq.(\ref{eq:prob_entang}) is displayed in Tab.\ref{tab:unica}, where it can be seen that $PCC=-1$ for $(q_\theta,q_\varphi)$ and also approximately for $(\tilde{\pi}_\varphi, \pi_\theta)$.

\begin{table}[h]
\centering
\begin{tabular}{|c|c|c|c|c|}
\hline
PCC(x,y) & $q_\theta$ & $\pi_\theta$ & $\tilde{\pi}_\varphi$ & $q_\varphi$ \\ \hline
$q_\theta$ & 1 & 0 & $-\beta$ & -1 \\ \cline{1-5}
$\pi_\theta$ & 0 & 1 & -1+$\beta^2/2$ & 0 \\ \hline
$\tilde{\pi}_\varphi$ & $-\beta$ & -1+$\beta^2/2$ & 1 & $\beta$ \\ \hline
$q_\varphi$ & -1 & 0 & $\beta$ & 1 \\ \hline
\end{tabular}
\caption{Pearson correlation coefficient matrix for the state in Eq. (\ref{eq:prob_entang}). The negative antidiagonal entries correspond to anti-correlations between quadratures.\label{tab:unica}}
\end{table}
In the limit $\beta\to 0 $, there is perfect anticorrelation between the azimuthal and polar angles as well as between their momenta. The meaning of the correlations in Tab. \ref{tab:unica} is that when the system reaches the state in Eq.(\ref{eq:prob_entang}), the measurement of $\theta$ will return a random value, but once $\theta$ is measured, $\varphi$ becomes strictly defined. For the particular state in Eq.(\ref{eq:prob_entang}), any value in $]-d/2, d/2[$ can be found equiprobably when measuring $q_\theta$ at $\zeta_E$. If the random outcome of measuring $q_\theta$ is $Q$, then the outcome of measuring $q_\varphi$ afterwards is $-\sqrt{2}Q$ with absolute certainty, and so $q_\theta$ and $q_\varphi$ fulfill the entanglement condition. The anti-correlations displayed in Tab. \ref{tab:unica} are the experimentally measurable quantities characterising entanglement, and they are analogous to the experimental entanglement measurements of \cite{palomakias}. Therefore, entanglement can be experimentally observed in liquid crystals as well.

Entanglement between the director angles arises from the fact that the probability distribution $\PP(\theta,\phi)$ describing the state of the system cannot be factorised into independent distributions for each director angle, which is due to the interaction between them. Bear in mind that if the state of the system is entangled, measuring one of the director's angles affects the measurement outcomes of its counterpart, but the change in the outcome probability as a consequence of the measurement does not imply that a real change has taken place in the system, only that one has gained additional information. Once the measurement is performed, there is a refinement of the knowledge over the state of the system, hence the correlations. The nature of entanglement becomes more tangible and clear with a frequentist view of probability theory. Consider the intuitive view of the probability of finding the director at a given orientation to be the relative frequency of the molecular orientation in the thermodynamic limit. Then the entanglement condition corresponds to the orientation of each molecule being random, but the difference between the polar and azimuthal angles being fixed and equal for every molecule.

Entanglement has also been experimentally tested under the fashionable form of inequality violations. Some of these violations \cite{palomakias,variaveiscontinuasnaoclassicas,variaveiscontinuasteorinfo,zoupereira} are based on the premise that if the product \cite{epreid} or sum \cite{inseparaveis,Duan} of the variances of entangled variables is below ZPM, then the variables must be entangled. It also assumes that the entanglement criterion requires a negative quasi-probability distribution \cite{inseparaveis}, and because of this negativity, the measured correlations must be nonclassical \cite{variaveiscontinuasnaoclassicas}. One of such inequalities reads \cite{Duan}:
\be
S=Var(q_1+q_2)+Var(p_1-p_2)\geq 2\,.
\label{eq:duanices}
\ee
For the state in Eq.(\ref{eq:prob_entang}) and with $1\to \theta, 2\to\phi$,
\be
S=Var(q_\theta+q_\varphi)+Var(\pi_\theta-\tilde{\pi}_\varphi)=\frac{3-\sqrt{8}}{12}d^2 + \left( \frac{3+\sqrt{8}}{4} + \frac{\beta^2}{8} \right)\sigma^2\,,
\label{eq:biolarei}
\ee
 and for $\delta,\sigma < 1$ the inequality can be violated. Thus, not only the inequality can be violated for a system without a quantum nature, it is also violated for a pair of  disentangled quadratures, such as  ($q_\theta+q_\phi,p_\theta-p_\phi$). As these inequalities are derived solely from a quantum formalism, they leave plenty of room for classical violations. What enables the inequality to be violated in the liquid crystal situation is the absence of restrictions in the spatial distribution of the director with regard to its orientation. The meaning of Eq.(\ref{eq:duanices}) is that the sum of the variances of $q_1+q_2$ and $p_1-p_2$ reaches below the zero-point uncertainty, but since there is no minimum uncertainty requirement for the molecular orientation, inequalities of this kind can be violated with low variance states, and even non-entangled quadratures. The advantage of the system presented here is that even if the variance at the boundary is large, it can be reduced by the squeezing effect of the cholesteric pitch and allow for the observation of violation of entanglement inequalities in liquid crystals.

\section{Validity and requirements}
\label{sec:validade}

A few requirements must be met for the above formalism to be valid. First, the validity of continuum elastic theory relies on $k\ll 2\pi/l_{mol}$, where $k$ is a macroscopic director wavelength and $l_{mol}$ the typical molecular size. In the strong-coupling regime considered, there are 2 distinct wavelengths: $\sqrt{2}\eta\lambda$ and $\lambda/(\sqrt{2}\eta)$. The short wavelength imposes the condition $P_0\gg \sqrt{8}K_2/(K_2+K_3) l_{mol}$. As all elastic moduli are of the same order of magnitude, the typical values of 300nm for the cholesteric pitch $P_0$ \cite{deGennes} and 3nm for the molecular size $l_{mol}$ guarantee that the condition is automatically satisfied. The long wavelength condition puts an upper limit on the intensity of the magnetic field $B\ll 2\pi \mu_0\sqrt[4]{8}\sqrt{K_2/(P_0l_{mol}\chi_a)}$. For the typical values of $K\sim 10^{-11}$N and $\chi_a\sim10^{-6}\mu_0$ \cite{lavrentovich, valoresCL} and using the same values for pitch and molecular size as before, the magnetic field upper threshold is at $\sim 1000$T. 

Another condition is that the free energy in Eq.(\ref{eq:energialivre}) is only valid for weakly twisted cholesterics, which means that the director gradient must be much smaller than the pitch wavelength. This roughly translates to $\lambda\ll \pi/P_0$, which places a second upper bound on the magnitude of the magnetic field: $B\ll \pi\mu_0/P_0 \sqrt{(K_2+K_3)/\chi_a}$, which for the values considered above means $B\ll 40$T.

In the previous sections, we restricted ourselves to the strong coupling regime $\eta \gg \sqrt{2}$, which places another bound on the magnetic field: $B\ll 4\pi \mu_0 K_2/(P_0\sqrt{2\chi_a(K_2+K_3)})$. For the values used above, this third upper bound reads
$B\ll 100$T. None of these upper limits invalidates the approximations used nor does it constitute any experimental limitation.

A fourth condition is that the system reaches the points $\zeta_T,\zeta_S,\zeta_E$ defined above, which means that the system size must be at least of the order of magnitude of the longest wavelength. This can be achieved by either enlarging the system or increasing the magnetic field, fulfilling the condition $B^2 L\gg 8\sqrt{2}\pi^2 \mu_0^2K_2/(P_0\chi_a)$, where $L$ is the system size. Translating this condition in terms of a lower bound for the magnetic field, and considering $L\sim 1$cm, one finds $B>2$mT, which is perfectly achievable with current technology. 
 
At last, we considered that the deviations from the potential minimum were small. Comparing the next order terms in the expansion of the potential in Eq.(\ref{eq:energialivre}), we find that the validity is assured for angular deviations much smaller than $80^\circ$, which is a considerable large range. Contrary to the dynamic case, when the system goes beyond the linear regime of Eq.(\ref{eq:Fquad}), the nonlinear spatial evolution does not trigger exponential amplifications, and so deviations from the potential minimum do not lead to drastic changes. 

Apart from the parameter regime required, it is also necessary to determine the local director field. The local director can be measured via the dependence on the incidence angle of light in reflectivity measurements \cite{refleteLC}, and recently, the local director field was also measured directly with a resolution up to the 100nm using scanning electron microscopy \cite{medirdiretor}. Therefore, we do not foresee any hindrance in the experimental observation of these effects in liquid crystals.

\section{Conclusion}

To conclude, a statistical framework was used to show that effects common in quantum systems can also be observed in cholesteric liquid crystals under the influence of an applied magnetic field. Particularly, the analysis of the statistical properties showed that state-swap, squeezing and entanglement naturally occur in liquid crystals as a consequence of the interaction. The formalism presented also enables the comparison to their quantum counterparts. Further, the absence of temporal dynamics (the system is in equilibrium and the spatial evolution plays the role of the time-evolution) in this liquid crystal scenario shields the system from the dissipation afflicting the dynamical case, which eases the observation of these effects. The discussion on the validity and feasibility of this proposal showed that the observation of these effects should be possible with current technology. Therefore, liquid crystals constitute a prime paradigm for a classical comparison to the physics displayed by many current devices probing quantum mechanics at a macroscopic scale \cite{sokawai,amiradosarabesunidos,palomakias,statetransfer}.

At last, because this liquid crystal scenario is a many-particles system in equilibrium and there is no actual dynamics, standard quantisation techniques are not immediately applicable. Direct quantisation attempts will even lead to a dimensional mismatch when introducing $\hbar$. Thus, state-swap, squeezing and entanglement are not quantum \textit{an sich}, and can occur in systems far from the quantum regime. Consequently, quantum claims based on these effects are, at best, a linguistic embellishment of the \textit{facta bruta}, where the term ``quantum'' is but a redundant definite article.

\ack
We thank C. Sch\"{a}fermeier and A. Silva for their useful comments, and the Dutch Science Foundation (NWO/FOM) for its financial support.

\appendix

\section{Proof of preservation of normalisation and positivity}
\label{app:provas}

For $\PP$ to represent a probability distribution governed by $d_t\PP=-\{\PP,\hml\}$, the following properties must be satisfied: (1) the probability distribution remains normalised; (2) the probability distribution remains positive. 

$\bullet$ To prove (1), let $\mathcal{N}(t)$ be the normalisation of the probability distribution. The time-evolution of $\mathcal{N}(t)$ is
\ba
d_t\mathcal{N}(t)&=\int dP dQ \,d_t\PP(Q,P,t)=\int dP dQ\, \{\PP,\hml\}\nonumber\\
&=-\int dP dQ\, \PP(\dpp_Q\dpp_P\hml-\dpp_P\dpp_Q\hml)=0
\ea
where it was used that $\PP$ must vanish at $Q,P=\infty$.
If $d_t\mathcal{N}(t)=0$ and if $\mathcal{N}(t=0)=1$, then the probability distribution is normalized at all times.

$\bullet$ To prove (2), note that for the time-evolution to be governed by the Poisson bracket, $\PP$ must be differentiable in all variables. Consequently, for a positive $\PP$ to become negative at a given point in phase-space, it must reach $0$ before it can take negative values. Let $(Q_0,P_0)$ be a point for which at $t=0$, the probability distribution is nonnegative everywhere and $\PP(Q_0,P_0,0)=0$. Then $(Q_0,P_0)$ is a minimum, and $\dpp_P\PP|_{(Q_0,P_0,0)}=\dpp_Q\PP|_{(Q_0,P_0,0)}=0$. Thus, for an arbitrarily small time-interval,
\ba
\PP (Q_0,P_0,\delta t)=\frac{1}{2}d_t^2\PP\bigg|_{(Q_0,P_0,0)}(\delta t)^2+O\big((\delta t)^3\big)\nonumber\\
\fl
=\Big(\dpp_Q^2\PP (\dpp_P\hml)^2+\dpp_P^2\PP(\dpp_Q\hml)^2
\quad -2(\dpp_P\dpp_Q\PP) (\dpp_P \hml)(\dpp_Q \hml)\Big)\bigg|_{(Q_0,P_0,0)}(\delta t)^2+O\big((\delta t)^3\big)
\ea
Since $(Q_0,P_0)$ is a minimum, $(\dpp_Q^2\PP, \dpp_P^2\PP)\geq0$, the Hessian matrix at $(Q_0,P_0)$ is positive semidefinite, and $(\dpp_Q^2\PP)(\dpp_P^2\PP)\geq (\dpp_P\dpp_Q\PP)^2$. Therefore,
\ba
\fl
\dpp_Q^2\PP (\dpp_P\hml)^2+\dpp_P^2\PP(\dpp_Q\hml)^2-2(\dpp_P\dpp_Q\PP) (\dpp_P \hml)(\dpp_Q \hml)\geq \nonumber \\
\geq\dpp_Q^2\PP (\dpp_P\hml)^2+\dpp_P^2\PP(\dpp_Q\hml)^2-2|\dpp_P\dpp_Q\PP|.|\dpp_P \hml|.|\dpp_Q \hml| \nonumber \\
\geq|\dpp_Q^2\PP|. |\dpp_P\hml|^2+|\dpp_P^2\PP|.|\dpp_Q\hml|^2 -2\sqrt{|\dpp_Q^2\PP|.|\dpp_P^2\PP|} |\dpp_P \hml||\dpp_Q \hml|\nonumber\\
\geq \Big(|\dpp_P \hml|\sqrt{|\dpp_Q^2\PP|}-|\dpp_Q \hml|\sqrt{|\dpp_P^2\PP|}\Big)^2\geq 0
\ea
Thus, the value of $\PP$ at any point $(Q_0,P_0)$ for which $\PP (Q_0,P_0)=0$ can never decrease, and so the probability distribution remains positive.\\

\section{Solution to the statistical dynamics}
\label{app:dinamica}

Using the method of characteristics to solve Eq.(\ref{eq:stocastica}), one arrives at the set of equations
\be
\dpp_\zeta \pi_\varphi=2q_\varphi
\ee
\be
\dpp_\zeta \pi_\theta = 4(1-2\eta^2)q_\theta - 4\eta^2 + 2\eta\pi_\varphi
\ee
plus the equations defining the momenta, Eqs.(\ref{eq:pi_teta},\ref{eq:pi_fi}). With $\chi_\pm=\sqrt{1-\eta^2\pm\eta\sqrt{\eta^2-2}}$, and including the chiral shift to polar momentum $\tilde{\pi}_\varphi:=\pi_\varphi-2\eta$, the solutions to this linear system of equations are

\ba
\fl
\sqrt{\eta^2-2}\,q_\varphi(\zeta) =& \frac{i}{\sqrt{2}}(\chi_-sgn(\sqrt{2}-\eta)\cosh (\chi_+\zeta)+\chi_+\cosh(\chi_-\zeta))q_\varphi(0)\nonumber\\
& + (-\chi_+\sinh (\chi_+\zeta)+\chi_-\sinh(\chi_-\zeta))q_\theta(0)\nonumber\\
 &+ \frac{1}{4}(-\cosh (\chi_+\zeta)+\cosh(\chi_-\zeta))\pi_\theta(0)\nonumber\\
 &+\frac{i}{\sqrt{8}}(sgn(\sqrt{2}-\eta)\sinh(\chi_+\zeta)+\sinh(\chi_-\zeta))\tilde{\pi}_\varphi(0)
\ea

\ba
\fl
\sqrt{\eta^2-2}\,q_\theta(\zeta) =& \frac{1}{2}sgn(\sqrt{2}-\eta)(\chi_-\sinh (\chi_+\zeta)-\chi_+\sinh(\chi_-\zeta))q_\varphi(0)\nonumber\\
& + \frac{i}{\sqrt{2}}(\chi_+\cosh (\chi_+\zeta)+sgn(\sqrt{2}-\eta)\chi_-\cosh(\chi_-\zeta))q_\theta(0)\nonumber\\
 &+ \frac{i}{4\sqrt{2}}(\sinh (\chi_+\zeta)+sgn(\sqrt{2}-\eta)\sinh(\chi_-\zeta))\pi_\theta(0)\nonumber\\
 &+\frac{1}{4}sgn(\sqrt{2}-\eta)(\cosh(\chi_+\zeta)-\cosh(\chi_-\zeta))\tilde{\pi}_\varphi(0)
\ea

\ba
\fl
\sqrt{\eta^2-2}\,\pi_\theta(\zeta) =& 2sgn(\sqrt{2}-\eta)(\cosh (\chi_+\zeta)-\cosh(\chi_-\zeta))q_\varphi(0) \nonumber\\
&+ i\sqrt{8}(\chi_+^2\sinh (\chi_+\zeta)+sgn(\sqrt{2}-\eta)\chi_-^2\sinh(\chi_-\zeta))q_\theta(0)\nonumber\\
 &+ \frac{i}{\sqrt{2}}(\chi_+\cosh (\chi_+\zeta)+sgn(\sqrt{2}-\eta)\chi_-\cosh(\chi_-\zeta))\pi_\theta(0)\nonumber\\
 &+sgn(\sqrt{2}-\eta)(\chi_+\sinh(\chi_+\zeta)-\chi_-\sinh(\chi_-\zeta))\tilde{\pi}_\varphi(0)
\ea

\ba
\fl
\sqrt{\eta^2-2}\,\tilde{\pi}_\varphi(\zeta) =& i\sqrt{2}(sgn(\sqrt{2}-\eta)\chi_-^2\sinh (\chi_+\zeta)+\chi_+^2\sinh(\chi_-\zeta))q_\varphi(0) \nonumber\\
&+ 2(-\cosh (\chi_+\zeta)+\cosh(\chi_-\zeta))q_\theta(0)\nonumber\\
 &+ \frac{1}{2}(-\chi_-\sinh (\chi_+\zeta)+\chi_+\sinh(\chi_-\zeta))\pi_\theta(0)\nonumber\\
 &+\frac{i}{\sqrt{2}}(sgn(\sqrt{2}-\eta)\chi_-\cosh(\chi_+\zeta)+\chi_+\cosh(\chi_-\zeta))\tilde{\pi}_\varphi(0)
\ea

In the strong-coupling regime ($\eta\gg 2$), $\chi_+\approx i/(\eta\sqrt{2})$, $\chi_-\approx i \eta \sqrt{2}$, and
\ba
\fl
q_\varphi(\zeta) \approx & \cos \left(\frac{\zeta}{\sqrt{2}\eta}\right)q_\varphi(0) -\sqrt{2}\sin(\sqrt{2}\eta\zeta)q_\theta(0)
+\frac{1}{4\eta}\left(\cos(\sqrt{2}\eta\zeta)-\cos \left(\frac{\zeta}{\sqrt{2}\eta}\right)\right)\pi_\theta(0)\nonumber\\
&+\frac{1}{\sqrt{8}\eta}\left(\sin \left(\frac{\zeta}{\sqrt{2}\eta}\right) - \sin(\sqrt{2}\eta\zeta)\right)\tilde{\pi}_\varphi(0)
\label{eq:q_fi_ap}
\ea

\ba
\fl
q_\theta(\zeta) \approx &\frac{1}{\sqrt{2}}\sin \left(\frac{\zeta}{\sqrt{2}\eta}\right) q_\varphi(0) + \cos (\sqrt{2}\eta\zeta) q_\theta(0)
+\frac{1}{4\sqrt{2}\eta}\left(\sin(\sqrt{2}\eta\zeta)-\sin \left(\frac{\zeta}{\sqrt{2}\eta}\right)\right)\pi_\theta(0)\nonumber\\
&+\frac{1}{4\eta}\left(\cos(\sqrt{2}\eta\zeta)-\cos\left(\frac{\zeta}{\sqrt{2}\eta}\right)\right)\tilde{\pi}_\varphi(0)
\ea

\ba
\fl
\pi_\theta(\zeta)\approx& \frac{2}{\eta}\left(\cos(\sqrt{2}\eta\zeta)-\cos \left(\frac{\zeta}{\sqrt{2}\eta}\right)\right)q_\varphi(0)-4\sqrt{2}\eta\sin(\sqrt{2}\eta\zeta)q_\theta(0)\nonumber\\
&+\cos(\sqrt{2}\eta\zeta)\pi_\theta(0)-\sqrt{2}\sin(\sqrt{2}\eta\zeta)\tilde{\pi}_\varphi(0)
\ea

\ba
\fl
\tilde{\pi}_\varphi(\zeta)\approx& -\sqrt{8}\eta\sin\left(\frac{\zeta}{\sqrt{2}\eta}\right)q_\varphi(0) + \frac{2}{\eta}\left(\cos(\sqrt{2}\eta\zeta)-\cos \left(\frac{\zeta}{\sqrt{2}\eta}\right)\right)q_\theta(0)\nonumber\\
&+\frac{1}{\sqrt{2}}\sin\left(\frac{\zeta}{\sqrt{2}\eta}\right)\pi_\theta(0)+\cos\left(\frac{\zeta}{\sqrt{2}\eta}\right)\tilde{\pi}_\varphi(0)
\label{eq:pi_fi_ap}
\ea

Considering that the probability distribution at the boundary factorises for each degree of freedom, the total probability distribution is given by
\be
\PP(\zeta)=\PP_{q_\varphi} (q_\varphi(0)) \PP_{q_\theta}(q_\theta(0)) \PP_{\pi_\theta}(\pi_\theta(0)) \PP_{\tilde{\pi}_\varphi}(\tilde{\pi}_\varphi(0))
\ee
where $\PP_j$ is the probability distribution at the boundary $\zeta=0$ of degree of freedom $j$. Inverting Eqs.(\ref{eq:q_fi_ap}-\ref{eq:pi_fi_ap}) to obtain the spatial evolution in terms of the phase space variables, and integrating in the momenta, one arrives at Eq.(\ref{eq:marg_prob}).

These solutions are valid everywhere except at the transition point between the weak-coupling and strong-coupling regimes $\eta=\sqrt{2}$. At this coupling point, the solutions to the linear system of equations given by the method of characteristics are

\ba
\fl
q_\varphi(\zeta)=& (\cos (\zeta) + \zeta \sin(\zeta))q_\varphi(0) - \sqrt{2}(\sin (\zeta) + \zeta \cos(\zeta))q_\theta(0) +\frac{\zeta}{2}\cos(\zeta) \pi_\varphi(0)\nonumber\\
&-\frac{\zeta}{\sqrt{8}}\sin(\zeta) \pi_\theta(0) - \sqrt{2}\zeta\cos (\zeta)
\ea

\ba
\fl
q_\theta(\zeta)=& (\cos (\zeta) - \zeta \sin(\zeta))q_\theta(0) +\frac{1}{\sqrt{2}}(\sin (\zeta) - \zeta \cos(\zeta))q_\varphi(0) +\frac{\zeta}{4}\cos(\zeta) \pi_\theta(0)\nonumber\\
& + \frac{\zeta}{\sqrt{8}}\sin(\zeta) \pi_\varphi(0) - \zeta\sin (\zeta)
\ea

\ba
\fl
\pi_\theta(\zeta)=& (\cos (\zeta) - \zeta \sin(\zeta))\pi_\theta(0) +\sqrt{2}(\sin (\zeta) + \zeta \cos(\zeta))\pi_\varphi(0) - 4(\sin (\zeta)+\zeta\cos(\zeta)) \nonumber\\
&-4(2\sin(\zeta)+\zeta \cos(\zeta)) q_\theta(0)+ \zeta\sin(\zeta) q_\varphi(0) 
\ea

\ba
\fl
\pi_\varphi(\zeta)=& (\cos (\zeta) + \zeta \sin(\zeta))\pi_\varphi(0) -\frac{1}{\sqrt{2}}(\sin (\zeta) - \zeta \cos(\zeta))\pi_\theta(0) - 4(\sin (\zeta)+\zeta\cos(\zeta)) \nonumber\\
&+2(2\sin(\zeta)-\zeta \cos(\zeta)) q_\varphi(0)-\sqrt{8}\zeta\sin(\zeta) q_\theta(0) 
\ea

\end{document}